\documentclass[aps,preprint,floats,nofootinbib]{revtex4}
\usepackage{amsmath}
\usepackage{amssymb}
\usepackage{latexsym}
\usepackage{psfrag}

\newlength{\defbaselineskip}
\newcommand{\setlinespacing}[1]%
           {\setlength{\baselineskip}{#1 \defbaselineskip}}

\usepackage{graphicx}
\usepackage{epsf}

\newcommand{\newc}{\newcommand}
\newcommand{\renewc}{\renewcommand}

\def\beq{\begin{equation}}
\def\eeq{\end{equation}}
\def\bea{\begin{eqnarray}}
\def\eea{\end{eqnarray}}
\def\besub{\begin{subequations}}
\def\eesub{\end{subequations}}
\newc{\ie}{{\it i.e. }}          \newc{\etal}{{\it et al. }}
\newc{\eg}{{\it e.g. }}          \newc{\etc}{{\it etc. }}
\newc{\cf}{{\it c.f. }}
\newc{\Tr}{\ensuremath{\mbox{Tr}}}
\newc{\mcL}{\mathcal{L}}
\newc{\mcE}{\mathcal{E}}
\newc{\mcQ}{\mathcal{Q}}
\newc{\mcU}{\mathcal{U}}
\newc{\mcD}{\mathcal{D}}

\newc{\gsim}{\lower.7ex\hbox{$\;\stackrel{\textstyle>}{\sim}\;$}}
\newc{\lsim}{\lower.7ex\hbox{$\;\stackrel{\textstyle<}{\sim}\;$}}

\renewc{\bar}{\overline}
\renewc{\P}{\ensuremath{\mbox{P}}}
\renewc{\d}{\ensuremath{\mbox{d}}}

\begin{document}
\thispagestyle{empty}

\preprint{
\hfill
\begin{minipage}[t]{3in}
\begin{flushright}
\vspace{0.0in}
OUTP-04/25P \\
FERMILAB-PUB-05-416-A
\end{flushright}
\end{minipage}
}

\hfill$\vcenter{\hbox{}}$

\vskip 0.5cm

\title{Improved Bounds on Universal Extra Dimensions
and Consequences for LKP Dark Matter}
\vskip 0.5cm
\author{\large{Thomas Flacke$^a$, Dan Hooper$^{b,c}$ and John March-Russell$^{a,d}$}}

\address{\vskip 0.5cm $^{a}$ Rudolf Peierls Center for Theoretical Physics, 1 Keble Road, Oxford University, Oxford OX1 3NP, UK \vskip 0.2cm\\
$^{b}$ Fermi National Accelerator Laboratory, Theoretical Astrophysics Center, Batavia, IL 60510, USA \vskip 0.2 cm\\
$^{c}$ Astrophysics Department, Oxford University, Oxford OX1 3RH, UK \vskip 0.2cm\\
$^{d}$ Department of Physics, University of California, Davis, CA  95616, USA\vskip 0.2cm}

\bigskip

\begin{abstract}

We study constraints on models with a flat ``Universal'' Extra 
Dimension in which all Standard Model fields propagate in the bulk. 
A significantly improved constraint on the compactification scale is obtained
from the extended set of electroweak precision observables accurately measured at LEP1 and LEP2. We find a lower bound of $M_c \equiv R^{-1} > 700$ (800) GeV
at the 99\% (95\%) confidence level. We also discuss the implications of thisconstraint on the prospects for the direct and indirect detection of Kaluza-Klein dark matter in this model.
\end{abstract}

\maketitle

\setcounter{page}{1}

\section{Introduction}

Models with flat ``Universal'' Extra Dimensions (UED) in which all
fields propagate in the extra dimensional bulk have received a much recent attention
\cite{Appelquist:2000nn} (for pre-dating ideas closely related to UED models see
Refs.\cite{preUED}).  They are of
phenomenological interest for two primary reasons.  First, among all
extra-dimensional models with Standard Model (SM) charged fields
propagating in the bulk, the mass scale of the compactification is most weakly
constrained \cite{Appelquist:2000nn},\cite{Appelquist:2002wb} - \cite{Buras:2003mk},
this mass scale being well within the reach of future collider experiments.  Moreover the collider
signatures of Kaluza-Klein (KK) particle production in UED models are easily confused with
those of superpartner production in some supersymmetric models \cite{fooled}.
Second, UED models provide a viable dark matter candidate -- the lightest
Kaluza-Klein particle (LKP) -- which is stable by virtue of a
conserved discrete quantum number intrinsic to the model.   Electroweak radiative
corrections imply that the LKP is neutral \cite{Cheng:2002iz} with a thermal relic density
consistent with observation for the mass range allowed by collider
bounds \cite{Cheng:2002ej,Servant:2002aq}.

Both of these features result from the existence of a conserved $Z_2$ KK parity.
Models in which all fields propagate in the extra dimensional
bulk allow conservation of a discrete (due to the finite volume) subgroup of
translation invariance, implying that the momentum in the extra dimension(s),
$p_i$, remains a conserved discrete quantity.  In terms
of the 4D effective theory, this translates into the conservation of
KK mode number.  However, in order to obtain chiral fermions in
the 4D effective theory, the extra dimension(s) have to be
compactified on an orbifold, which inevitably breaks translation invariance and
hence induces KK-number violation, but still preserves KK-parity
as a conserved $Z_2$ quantum number.  This KK-parity implies processes
with a single first KK excitation and Standard
Model particles only are forbidden, and the lightest KK-particle is
stable. 

Specializing to the case of one extra dimension, the bound on the
compactification scale $M_c\equiv 1/R$ from direct non-detection is
$M_c\gsim 300$~GeV \cite{Appelquist:2000nn,Agashe:2001xt} 
being a factor of 2 lower than the na\"{\i}ve estimate for models without
KK parity conservation.  Comparable
bounds have been derived from the analysis
of Electro-Weak Precision Tests (EWPT) 
\cite{Appelquist:2000nn, Appelquist:2002wb}, the improvement of which is
a primary focus of this paper.   The bound from the
$Z\rightarrow b\bar{b}$ branching ratio is of order $M_c\gsim 200$ GeV \cite{Appelquist:2000nn}
(the bounds from the $Z\rightarrow b\bar{b}$ left-right asymmetry, and from the muon anomalous
magenetic moment, are significantly weaker),
while $b\rightarrow s \gamma$ leads to $M_c\gsim 280$ GeV \cite{Agashe:2001xt} and 
constraints from FCNC processes \cite{Buras:2002ej,Buras:2003mk} are of the order 
of $M_c\gsim 250$ GeV.  Concerning current and future experiments, Run II at the Tevatron will be able to
detect KK-particles if $M_c\lsim 600$~GeV while the LHC will reach $M_c\sim 3$~TeV
\cite{Rizzo:2001sd,Macesanu:2002db}, well beyond the currently
excluded compactification scale.

In this paper, we will extend the analysis of constraints from EWPT
to arrive at a significantly more stringent bound on the compactification scale. Within
the remaining allowed parameter space, we then investigate the direct and indirect detection
prospects of the LKP dark matter candidate.

Specifically, in Section~2 we briefly review UED models, following
Refs.\cite{Appelquist:2000nn,Appelquist:2002wb}.
 In Sections 3 and 4, the
constraints from EWPT are investigated
including the full LEP2 data set following the general analysis of
\cite{Barbieri:2004qk}.  As has been shown in \cite{Barbieri:2004qk},
this {\it a priori} requires an extended set of EWPT parameters, and we
calculate the contributions from universal extra dimension models
to this extended set.  A fit to the LEP1 and LEP2 data set
leads to significantly improved bounds
on $M_c$ as a function of the unknown Higgs mass $m_H$.
We emphasize that the two-loop Standard Model Higgs
contributions to the EWPT parameters are included in this fit
following the simple accurate numerical interpolation of
\cite{Barbieri:2004qk}.   In Section~5 we discuss the implications of
this improved bound for KK dark matter, particularly the prospects of
direct and indirect detection.  Finally Section~6 contains our
conclusions while an appendix examines, in an improved version of na\"{\i}ve
dimensional analysis, the maximum scale of applicability
of the 5-dimensional UED theory with which we calculate.

\section{The UED Model}

We consider the 5-dimensional extension of the single Higgs doublet Standard Model
with all fields propagating in the extra dimension.  The 5D Lagrangian is
\bea\label{L5D}
\mathcal{L}_{5D}=&-&\frac{1}{4}G^A_{MN}G^{AMN}-\frac{1}{4}W^I_{MN}W^{IMN}-
\frac{1}{4}B_{MN}B^{MN}\nonumber\\
&+&(D_MH)^{\dagger}(D^MH)+\mu^2H^{\dagger}H-\frac{1}{2}\lambda(H^{\dagger}H)^2+i \bar{\psi} \gamma^M D_M \psi\\
&+&\left(\hat{\lambda}_{E}\bar{\mcL}\mcE H+\hat{\lambda}_{U}\bar{\mcQ}\mcU\tilde{H}+
\hat{\lambda}_{D}\bar{\mcQ}\mcD H+\mbox{h.c.}\right)+\ldots\nonumber
\eea
where $G_{MN}$, $W_{MN}$, $B_{MN}$ are the 5D $SU(3)_C\times SU(2)_W\times U(1)_Y$
gauge field strengths, the covariant
derivatives are defined as $D_M=\partial_M+i \hat{g}_3 G^A_{M}T^A+i
\hat{g}_2 W^I_{M}T^I+i \hat{g}_1 YB_M $, where $\hat{g}_i$
are the 5D gauge couplings, with engineering dimension $m^{-1/2}$.
The ellipses in Eq.(\ref{L5D}) denote higher-dimension  
operators whose effect we discuss below and in the appendix.
For compactification on $S^1$,  
the 5D matter fermions $\psi = (\mcQ,\mcU,\mcD,\mcL,\mcE)$ contain in 4D language
both left-handed and right-handed chirality zero modes, eg, $\mcQ=(Q_L,Q_R)$.
The 5D Higgs scalar $H$ is in the representation
$(1,2)_{1/2}$ and $\tilde{H}=i\sigma_2H^*$, and for simplicity the family
indicies on the fields and 5D Yukawa couplings, 
$\hat{\lambda}$, are suppressed.  

Chiral fermions are obtained at the KK zero mode level by
compactifying the extra dimension on the orbifold $S^1/Z_2$.
The length of the orbifold is $\pi R$ and the 
associated KK mass scale $M_c\equiv 1/R$.  By integrating out the extra dimension, 
every 5D field yields an infinite tower of effective 4D Kaluza-Klein 
modes. For compactification on $S^1$, the theory would be translation 
invariant in the extra dimension, implying conservation of 5-momentum and 
therefore KK mode number $k$ ($\sum_i k_i=0$) in every vertex, however, for 
compactification on an orbifold, the orbifold boundaries break translation 
invariance and therefore 5-momentum conservation. In the KK picture, this 
corresponds to the existence of KK number violating operators (which
are induced at loop-level even if not included in the bare 5D Lagrangian).
By definition of the $S^1/Z_2$ orbifold, the 5D theory however 
still has a symmetry under reflection in the extra dimension $y \rightarrow -y$. 
In terms of KK-modes this translates into the conservation of KK-parity: 
\beq
\sum_i k_i=0~\mbox{mod}~2. 
\eeq

Choosing even boundary conditions for all gauge fields and $H$ as well as the chiral
fermion components $Q_L$, $U_R$, $D_R$, $L_L$, and $E_R$
(but not their would-be mirror partners), the resulting KK-zero modes are identical to
the SM fields.  At tree-level, the parameters in Eq.(\ref{L5D}) are
defined in terms of the SM couplings and $R$ via $\hat{g}_i^2\equiv \pi R g_i^2$ and 
$\hat{\lambda}_i^2\equiv \pi R \lambda_i^2$. At tree-level the mass of the $j$-th 
KK-mode is given by $m^{(j)}_i=\sqrt{(M_c)^2+m^2_i}$ for fermions 
and gauge bosons with the first KK-excitation of the photon being the lightest KK-particle.
Due to KK-parity conservation, the LKP is stable, providing a dark matter candidate. 
The KK-interaction vertices for the UED model are
given in Refs.\cite{Appelquist:2000nn,Petriello:2002uu,Buras:2002ej}.

\section{Constraints from Precision Measurements at LEP1 and LEP2}\label{chconstraints}

Even below the KK-particle production threshold $\sim 2M_c$,
KK-particles enter experimental constraints via virtual effects 
in radiative corrections.  At one-loop level and at LEP
energies, vertex corrections and box diagrams are suppressed
compared to self-energy contributions \cite{Appelquist:2002wb}. 
Higgs-KK mode contributions to
vertices and box diagrams are suppressed by small Yukawa couplings
because no top pair can be produced at LEP energies, while
KK-gauge boson contributions are parametrically suppressed by a factor 
of $(m_W/M_c)^2$.   Thus the UED radiative corrections 
are approximately oblique, \ie\, flavor independent. Oblique corrections
are traditionally parameterized by the Peskin-Takeuchi $\hat{S},\hat{T},\hat{U}$
parameters \cite{Peskin:1990zt} or equivalently by the EWPT parameters
$\epsilon_{1,2,3}$ \cite{Altarelli:1990zd}, both of which are
defined in terms of the gauge-boson self-energies.

Using an effective field theory approach, it has been shown in
\cite{Barbieri:2004qk} that $\hat{S}, \hat{T}$, and $\hat{U}$ do
not form a complete parameterization of relevant corrections 
to the Standard Model, where the only significant deviations from the SM
reside in the self-energies of the vector bosons.  (We will call these
corrections {\it oblique} as opposed to {\it universal} as is 
done in \cite{Barbieri:2004qk} to avoid confusion.)  If the scale
of new physics is above LEP2 energies then the new physics
contributions to the transverse gauge boson self energies
are analytic in $q^2$ and can be power series expanded.
The full independent set of electroweak precision observables (EWPO)
that are well-determined by the LEP1 and LEP2 data sets can be
defined by\footnote{The observables defined in (\ref{STUVXYWdef}) differ by
factors of $g$ and $g'$ compared to \cite{Barbieri:2004qk} as we
employ canonical normalizations for the gauge bosons.}
\bea\label{STUVXYWdef}
\hat{T} &\equiv& \frac{1}{m_W^2}\left(\Pi_{W_3W_3}(0)-\Pi_{W^+W^-}(0)\right)\nonumber\\
\hat{S} &\equiv& \frac{g}{g'}\Pi'_{W_3B}(0)\nonumber\\
\hat{U} &\equiv& \Pi'_{W^+W^-}(0)-\Pi'_{W_3W_3}(0)\nonumber\\
X &\equiv& \frac{m_W^2}{2}\Pi''_{W_3B}(0)\\
Y &\equiv& \frac{m_W^2}{2}\Pi''_{BB}(0)\nonumber\\
W &\equiv& \frac{m_W^2}{2}\Pi''_{W_3W_3}(0)\nonumber
\eea
where $\Pi$ denote the new-physics contributions to the
transverse gauge boson vacuum polarization amplitudes,
with $\Pi'(0) = d\Pi(q^2)/dq^2 |_{q^2=0}$, etc.   {\it A priori}, four more parameters
are needed in addition to $S,T,U$ in order to parameterize the
full freedom in the electroweak gauge boson self energy corrections up to
order $(q^2)^2$, but only $X,Y$ and $W$ are well-determined by the LEP data
sets.

\begin{figure}[t]
\begin{center}
\includegraphics[width=4.0in]{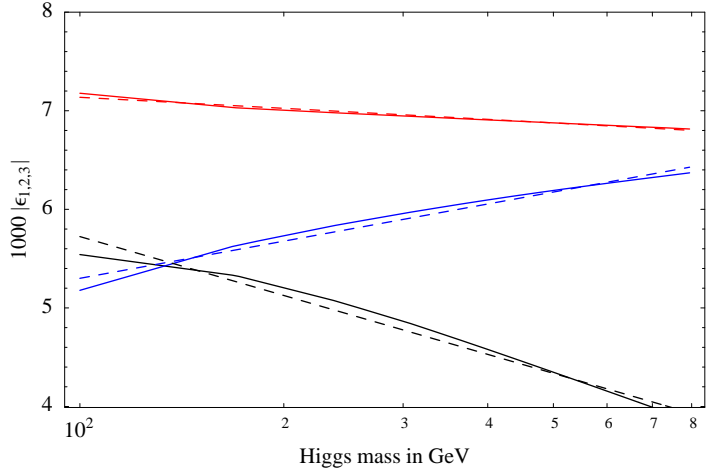}
\caption{Comparison of the 2-loop order Higgs-dependent contributions to the electroweak gauge
boson radiative corrections as implemented in the TopaZ0 code (solid lines) with the simple
numerical interpolations (dashed lines) given in Eq.(\ref{SMepsilon}).}
\label{interpolation}
\end{center}
\end{figure}

A convenient way of expressing the $Z$-pole LEP1 experimental constraints 
on the electroweak precision observables is in terms of the
$\epsilon_1,\epsilon_2,\epsilon_3$ parameters whose determination
are independent of the unknown mass of the Higgs. 
The experimental constraints from LEP1 determine:
\begin{equation}\label{epsilonexp}
    \begin{array} {ccc}
        \begin{array}{c}
            \epsilon_1=+(5.0\pm1.1)10^{-3}\\
            \epsilon_2=-(8.8\pm1.2)10^{-3}\\
            \epsilon_3=+(4.8\pm1.0)10^{-3}\\
        \end{array} &
        \begin{array}{c}
                                          \\
            \mbox{with correlation matrix}\\
                                          \\
        \end{array} &
                 \rho= \left(\begin{array}{ccc}
              1 & 0.66 & 0.88\\
              0.66 & 1 & 0.46\\
              0.88 & 0.46 & 1\\
        \end{array}\right)
    \end{array}
\end{equation}
These observables are related to
the $\hat{S},\hat{T},\hat{U},X,Y$ and $W$ parameters by
\cite{Barbieri:2004qk}
\bea\label{defepsilon}
\epsilon_1 &=& \epsilon_{1,SM}+\hat{T}-W+2X\frac{\sin\theta_W}{\cos\theta_W}
-Y\frac{\sin^2\theta_W}{\cos^2\theta_W}\nonumber\\
\epsilon_2 &=& \epsilon_{2,SM}+\hat{U}-W+2 X\frac{\sin\theta_W}{\cos\theta_W} \\
\epsilon_3 &=& \epsilon_{3,SM}+\hat{S}-W+\frac{X}{\sin\theta_W\cos\theta_W }-Y .\nonumber
\eea
where $\epsilon_{i,SM}$ denote the Higgs-dependent contributions to the electroweak gauge
boson radiative corrections, which by definition are not included in $\hat{S},\hat{T},\hat{U},X,Y$
and $W$. The full Higgs-dependent corrections $\epsilon_{i,SM}$ have been calculated to 2-loop order
\cite{Degrassi:1996ps,Degrassi:1996mg} and implemented in precision electroweak codes such as as TopaZ0 \cite{Montagna:1993ai,Montagna:1998kp}.
The authors of Ref.\cite{Barbieri:2004qk} give a simple but accurate numerical interpolation
to these full 2-loop order results, as shown in Figure \ref{interpolation}.\footnote{We thank the
authors of Ref.\cite{Barbieri:2004qk} for communications regarding this interpolation, and especially
Alessandro Strumia for generating Figure \ref{interpolation} for us.}
\bea\label{SMepsilon}
\epsilon_{1,SM}&=&\left(+6.0-0.86 \ln \frac{m_H}{m_Z}\right)10^{-3}\nonumber\\
\epsilon_{2,SM}&=&\left(-7.5+0.17 \ln \frac{m_H}{m_Z}\right)10^{-3}\\
\epsilon_{3,SM}&=&\left(+5.2+0.54 \ln \frac{m_H}{m_Z}\right)10^{-3}.\nonumber
\eea

Similarly the full LEP2 data set at center-of-mass energies 
varying from 189 GeV to 207 GeV lead to the following determination of the 
$X,Y,W$ parameters \cite{Barbieri:2004qk}
\begin{equation}\label{XYWexp}
    \begin{array} {ccc}
        \begin{array}{c}
            X=(-2.3\pm3.5)10^{-3}\\
            Y=(+4.2\pm4.9)10^{-3}\\
            W=(-2.7\pm2.0)10^{-3}\\
        \end{array} &
        \begin{array}{c}
                                          \\
            \mbox{with correlations}\\
                                          \\
        \end{array} &
                 \rho= \left(\begin{array}{ccc}
              1 & -0.96 & +0.84\\
              -0.96 & 1 & -0.92\\
              +0.84 & -0.92 & 1\\
        \end{array}\right).
    \end{array}
\end{equation}

The EWPO $\hat{S},\hat{T},Y$ and $W$ break different parts of $SU(2)_L\times
SU(2)_{\mbox{custodial}}$.  Without knowing the specific high
energy completion there is no physical reason for a hierarchy
between them whereas $\hat{U}$ and $X$ correspond to higher derivative
operators with the same symmetry properties as $\hat{T}$ and $\hat{S}$ respectively.
Thus if $M$ is the scale of new physics, the expectation is that new physics
contributions to $\hat{U}$ and $X$ will be suppressed by powers of
$(m_W/M)^2$ compared to the new physics contributions to
$\hat{S},\hat{T},Y$ and $W$.  However for the purpose of investigating the bounds on 
new physics, such as the KK mass scale $M_c$, it is not correct to restrict the analysis to
$\hat{S},\hat{T},Y,W$, and exclude $\hat{U}$ and $X$.   Even though the new physics contributions
to $\hat{U}$ and $X$ are expected to be
suppressed from an effective theory point of view, this is not
reflected in the experimental constraints on these EWPO. For example 
the LEP2 data determine $X=(-2.3\pm 3.5) 10^{-3}$ with a similar
accuracy as the ``dominant'' parameters. The fact that the experimentally
preferred value is {\it not} suppressed ought to be included in
the analysis.  We therefore keep the full parameter set.

\section{Constraints on UED Models from EWPO}\label{chEWPTanalysis}

In this section we calculate the contributions to the extended
set of EWPO from UED models and obtain an improved constraint
on $1/R$ by performing a $\chi^2$-fit to the experimental values
given in Eqs.(\ref{epsilonexp}-\ref{XYWexp}).  We note in passing that
the consequences of the combined LEP1 and LEP2
constraints have so far been explored in 5D models with gauge bosons
in the extra dimension and Higgsless models
\cite{Barbieri:2004qk}, for supersymmetry \cite{Marandella:2005wc}
and for Little Higgs Models \cite{Barbieri:2004qk,Marandella:2005wd}.

Concerning UED models, for low Higgs mass, the dominant constraint
on $1/R$ is expected from the measurement of $T$ rather than
$S$ while $U$ is further suppressed.  For large
$m_H$ however, the one-loop Standard Model contribution to $T$ can
compensate for the KK-contribution such that a combined $S,T$
analysis is necessary.  In \cite{Appelquist:2002wb}, one-loop
KK contributions to the $S$ and $T$ parameters are fitted to the experimentally
determined values of $S,T$ from LEP1, yielding a constraint on the $(1/R,m_H)$ parameter
space (see Figure 3 of Ref.\cite{Appelquist:2002wb}). The lowest allowed
compactification scale is $1/R\sim 300$ GeV at high $m_H\sim 800$
GeV as a result of this cancellation in the $T$ parameter.  Due to the
significant dependence of the current limit
on the cancellation in contributions to $T$ at large $m_H$ where higher
terms in the loop expansion for the Higgs-dependent contributions
are becoming large, it is important to include the two-loop
SM contributions to test if this cancellation is stable.   We show below
that the 2-loop Higgs contributions destroy the cancellation resulting
in an improved constraint on $M_c=1/R$.
A further improvement results from the inclusion of LEP2
data, specifically the measurements of the $X,Y,W$ parameters.

We have calculated the one-loop KK contributions to the full set of electro-weak
precision observables 
$\hat{S},\hat{T},\hat{U},X,Y$ and $W$.   These contributions,
which for one extra dimension $\delta=1$ are functions of $1/R$ and $m_H$
are in general extremely complicated and rather unilluminating in their explicit
form.  

\begin{figure}[t]
\psfrag{mH}[c]{$m_H$ [GeV]}
\psfrag{T}[c]{$\hat{T}_x$}
\begin{center}
\includegraphics[width=4.0in]{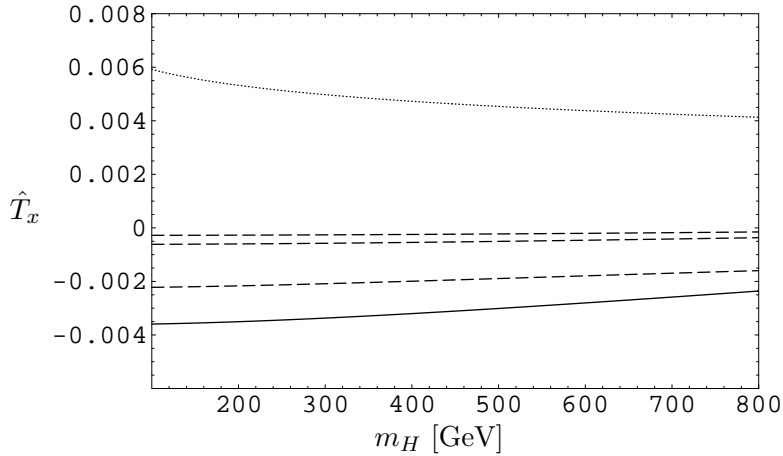}
\caption{The contribution to $\hat{T}$ from the first three KK levels (dashed lines) for $M_c=400$ GeV
as a function of Higgs mass in the range 100 to 800 GeV, as well as the sum over the first 10 KK modes
(solid line) and the numerically-interpolated Higgs-dependent correction (dotted line) arising from $\epsilon_{1,SM}$.}
\label{ttest}
\end{center}
\end{figure}
As an example of the KK contributions, Figure \ref{ttest} shows the Standard Model
contribution to $\epsilon_1$ and the first 
three KK-modes of $\hat{T}$ as well as the sum over the first 10 KK modes at $1/R=400$ GeV.
Similar behavior occurs for the other EWPO.  
In all cases the sum over KK-modes converges sufficiently fast
such that in our further analysis we approximate the UED contributions to the EWPO by the
sum over the first 10 modes. 

\begin{figure}[t]
\psfrag{mH}[c]{$m_H$  [GeV]}
\psfrag{Rinv}[c]{$R^{-1}$  [GeV]}
\begin{center}
\includegraphics[width=4.0in]{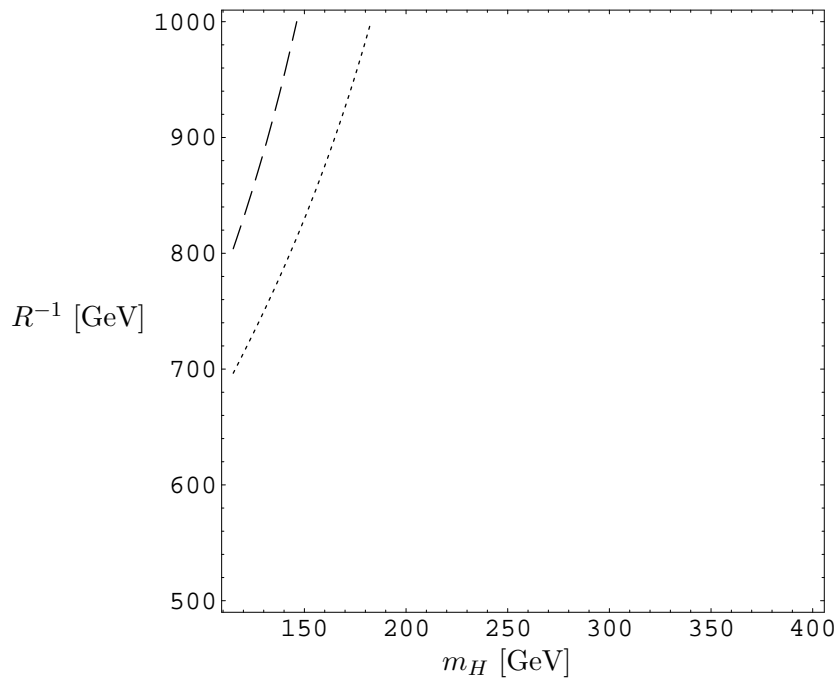}
\caption{The 95\% (dashed line) and 99\%  (dotted line) confidence limit exclusion zones for the UED model, as
a function of Higgs mass in the range 115 GeV to 400 GeV, and mass $M_1=1/R$ of the lightest
KK excitation in the range 500 GeV to 1 TeV. The excluded regions are towards the bottom right
of the figure.}
\label{chi9599}
\end{center}
\end{figure}

Using the expressions for $\hat{S},\hat{T},\hat{U},X,Y,W$ we have derived from the
UED model we have performed a $\chi^2$ fit to the LEP1 and LEP2 experimental data
encapsulated in Eqs.(\ref{epsilonexp}-\ref{XYWexp}).  Figure \ref{chi9599} shows the
resulting constraints on the $1/R$, $m_H$ parameter space. 

We find that the lower bound on the mass of the first KK level is improved
to $M_c\equiv R^{-1}>700 (800)$ GeV at the 99\% (95\%) confidence level.
There are two origins for the improvement of the bound compared to
\cite{Appelquist:2002wb}.  First, when taking two-loop Standard Model
contributions to the electroweak precision parameters into account, the
KK-contributions to the $\hat{T}$ parameter no longer cancel against
Higgs-dependent contributions in the heavy-Higgs-mass limit.  This
differs from the situation of Ref.\cite{Appelquist:2002wb} where only
one-loop Higgs-dependent contributions were taken in to account.  This
lack of cancellation is the reason for elimination of the heavy-Higgs-mass
and low $M_c$ region.   Second, the inclusion of LEP2
data into the analysis necessitated the use of an extended set of well-determined
electroweak precision observables as shown in Ref.\cite{Barbieri:2004qk}. 
These new EWPO provide additional constraints, further lifting the bound on $M_c$.

\section{KK Dark Matter}

In recent years, the Lightest KK Particle (LKP) in UED models has become a rather
popular candidate for the dark matter of our Universe \cite{Cheng:2002ej,Servant:2002aq}. In
this section, we will review the phenomenology of KK dark matter in this scenario, and discuss
the detection prospects for such a particle in light of the new electroweak precision
constraints presented in this article. 

As stated earlier, the most natural choice for the LKP in UED models is the first KK excitation
of the hypercharge gauge boson, $B^{(1)}$. Such a state, being electrically neutral and colorless,
can serve as a viable candidate for dark matter. One attractive feature of this candidate is
that its thermal relic abundance is naturally near the measured quantity of cold dark matter
for $M_c \equiv R^{-1} \sim$ TeV.

The number density of the LKP evolves according to the Boltzman equation
\begin{equation}
\frac{dn_{B^{(1)}}}{dt} + 3 H n_{B^{(1)}}
= -<\sigma v> \bigg[(n_{B^{(1)}})^2 - (n^{\rm{eq}}_{B^{(1)}})^2 \bigg], 
\end{equation}
where $H$ is the Hubble rate, $n^{\rm{eq}}_{B^{(1)}}$ denotes the equlibrium number
density of the LKP, and $<\sigma v>$ is the LKP's self-annihilation cross section,
given by~\footnote{The LKP self-annihilation cross section may be enhanced by processes
involving the resonant s-channel exchange of second level KK modes, particularly if the
mass of the higgs boson is somewhat large \cite{kkreson}.}
\begin{equation}
<\sigma v> \simeq \frac{95 g^4_1}{324 \pi m^2_{B^{(1)}}}.
\label{annlkp}
\end{equation}
Numerical solutions of the Boltzman equation yield a relic density of 
\begin{equation}
\Omega_{B^{(1)}} h^2 \approx \frac{1.04 \times 10^9 x_F}{M_{\rm{Pl}} \sqrt{g^*} (a+3b/x_F)},
\label{sol}
\end{equation}
where $x_F = m_{B^{(1)}}/T_F$, $T_F$ is the relic freeze-out temperature, $g^*$ is
the number of relativistic degrees of freedom available at freeze out ($g^* \simeq 92$
for the case at hand), and $a$ and $b$ are terms in the partial wave expansion of the
annihilation cross section, $\sigma v = a + bv^2 +\vartheta(v^4)$. Note that in this
simple case with only the LKP participating in the freeze-out process, $b$ can safely
be neglected. Evaluation of $x_F$ leads to
\begin{equation}
x_F = \ln\bigg[c(c+2) \sqrt{\frac{45}{8}}
\frac{g m_{B^{(1)}} M_{\rm{Pl}}  (a+6b/x_F)}{ 2 \pi^3 \sqrt{g^* x_F}}\bigg],
\label{xf}
\end{equation}
where $c$ is an order 1 parameter determined numerically and $g$ is the number of
degrees of freedom of the LKP. Note that since $x_F$ appears in the logarithm as
well as on the left hand side of the equation, this expression must be solved by
iteration. WIMPs generically freeze-out at temperatures in the range of approximately
$x_F\approx$ 20 to 30.

When the cross section of Eq.(\ref{annlkp}) is inserted into Eqs.(\ref{sol}) and~(\ref{xf}),
a relic abundance within the range of the cold dark matter density measured by WMAP
($0.095 < \Omega h^2 < 0.129$) \cite{wmap} can be attained for $m_{B^{(1)}}$ approximately
in the range of 850 to 950 GeV. This conclusion can be substantially modified if other KK
modes contribute to the freeze-out process, however.

To include the effects of other KK modes in the freeze-out process, we adopt the
following formalism. In Eqs.(\ref{sol}) and~(\ref{xf}), we replace the cross section
($\sigma$, denoting the appropriate combinations of $a$ and $b$) with an effective
quantity which accounts for all particle species involved
\begin{equation}
\sigma_{\rm{eff}} = \sum_{i,j} \sigma_{i,j} \frac{g_i g_j}{g^2_{\rm{eff}}}
\, (1+\Delta_i)^{3/2} \, (1+\Delta_j)^{3/2} \, e^{-x(\Delta_i+\Delta_j)}. 
\end{equation}
Similarly, we replace the number of degrees of freedom, $g$, with the effective quantity
\begin{equation}
g_{\rm{eff}} = \sum_{i} g_i \, (1+\Delta_i)^{3/2} e^{-x \Delta_i}.
\end{equation}
In these expressions, the sums are over KK species, $\sigma_{i,j}$ denotes the
coannihilation cross section between species $i$ and $j$ and the $\Delta$'s
denote the fractional mass splitting between that state and the LKP.

To illustrate how the presence of multiple KK species can affect the freeze-out
process, we will describe two example cases. First, consider a case in which the
coannihilation cross section between the two species, $\sigma_{1,2}$, is large
compared to the LKP's self-annihilation cross section, $\sigma_{1,1}$. If the
second state is not much heavier than the LKP ($\Delta_2$ is small), then
$\sigma_{\rm{eff}}$ may be considerably larger than $\sigma_{1,1}$, and thus
the residual relic density of the LKP will be reduced. Physically, this case
represents a second particle species depleting the WIMP's density through
coannihilations. This effect is often found in the case of supersymmetry models
in which coannihilations between the lightest neutralino and another superpartner,
such as a chargino, stau, stop, gluino or heavier neutralino, can substantially
reduce the abundance of neutralino dark matter.

The second illustrative case is quite different. If $\sigma_{1,2}$ is comparatively
small, then the effective cross section tends toward $\sigma_{\rm{eff}} \approx
\sigma_{1,1}\, g^2_1/(g_1+g_2)^2 + \sigma_{2,2}\, g^2_2/(g_1+g_2)^2$. If $\sigma_{2,2}$
is not too large, $\sigma_{\rm{eff}}$ may be smaller than the LKP's self-annihilation
cross section alone. Physically, this scenario corresponds to two species freezing out
quasi-independently, followed by the heavier species decaying into the LKP, thus
enhancing its relic density. Although this second case does not often apply to
neutralinos, KK dark matter particles may behave in this way for some possible
arrangements of the KK spectra.

Very recently, the LKP freeze-out calculation has been performed, including all
coannihilation channels, by two independent groups~\cite{coannkribs,coannmat}. We will
summarize their conclusions briefly here. 

As expected, the effects of coannihilation on the LKP relic abundance depend critically
on the KK spectrum considered. If strongly interacting KK states are less than roughly
$\sim 10\%$ more heavy than the LKP mass, the effective LKP annihilation cross section
can be considerably enhanced, thus reducing the relic abundance. KK quarks which are
between 5\% and 1\% more massive than the LKP lead to an LKP with the measured dark
matter abundance over the range of masses, $m_{B^{(1)}} \approx $ 1500 to 2000 GeV.
If KK gluons are also present with similar masses, $m_{B^{(1)}}$ as heavy as 2100 to
2700 GeV is required to generate the observed relic abundance. We thus conclude that
if KK quarks or KK gluons are not much more massive than the LKP, the new constraints
presented in this article do not reach the mass range consistent with the observed
abundance of cold dark matter.

On the other hand, it is possible that all of the strongly interacting KK modes may be
considerably more heavy than the LKP. In this circumstance, other KK states may still
affect the LKP's relic abundance. If, for example, all three families of KK leptons are
each 1\% more massive than the LKP, the observed relic abundance is generated only for
$m_{B^{(1)}}$ between approximately 550 and 650 GeV. This range is excluded by the constraints
presented in this article. If the KK leptons are instead 5\% more massive than the LKP, the
observed abundance is found for $m_{B^{(1)}}\approx$ 670 to 730 GeV, which is excluded at around the 99\% confidence level. We thus conclude that electroweak precision measurements
are not consistent with dark matter in this model if KK leptons are within approximately 5\%
of the LKP mass, unless other KK states are also quasi-degenerate.

The constraints put forth here can have a substantial impact on the prospects for the direct
and indirect detection of KK dark matter. Firstly, direct detection experiments benefit
from the larger elastic scattering cross sections found for smaller values of $m_{B^{(1)}}$.
Spin-dependent scattering of the LKP scales with $1/m^4_{B^{(1)}}$, while the spin-independent
cross section goes like $1/(m^2_{B^{(1)}}-m^2_{q^{(1)}})^2$ \cite{elastic}. In either case,
the largest scattering rates are expected for the lightest LKPs. Furthermore, heavier WIMPs
have a smaller local number density, and thus a smaller scattering rate in direct detection
experiments.

Even if the new constraints presented here are not taken into account, the prospects for the
direct detection of KK dark matter is somewhat poor. Cross sections are expected to be smaller
than roughly $10^{-9}$ pb and $10^{-4}$ pb for spin-independent and dependent scattering,
respectively \cite{elastic}, both of which are well beyond the reach of existing experiments.
Next generation experiments may be able to reach this level of sensitivity, however.

The situation is rather different for the case of indirect detection. The annihilation
cross section of LKPs is proportional to $1/m^2_{B^{(1)}}$, and thus a heavier LKP corresponds
to a lower rate of annihilation products being generated in regions such as the galactic center,
the local halo, external galaxies and in local substructure. For dark matter searches using
gamma-rays from regions such as the galactic center \cite{gamma}, this is probably of marginal
consequence, considering the very large astrophysical uncertainties involved. The annihilation
rate of LKPs in the local halo, however, has less associated astrophysical uncertainty. LKP
annihilations in the galactic halo to anti-matter particles \cite{antimatter}, positrons in
particular, may be potentially observable if the LKP annihilation cross section is large enough,
{\it i.e}.~if the LKP is sufficiently light. It has been shown \cite{poskribs} that the
cosmic positron excess observed by the HEAT experiment \cite{heat} could have been generated
through the annihilations of LKPs in the surrounding few kiloparsecs of our galaxy.  This, however,
requires $m_{B^{(1)}}$ to be in the range of approximately 300 to 400 GeV \footnote{A large
degree of local dark matter substructure, if present, may potentially enable this mass range
to be extended to considerably higher values.}, which is strongly excluded by the results
presented in this article. Even if the HEAT excess is not a product of dark matter annihilations,
the presence of KK dark matter in the local halo will possibly be within the reach of future
cosmic positron measurements, particularly those of the AMS-02 experiment \cite{ams}. Assuming
a modest degree of local inhomogenity, LKP masses up to $m_{B^{(1)}} \approx 900$ GeV should be
within the reach of AMS-02 \cite{possilk}.

Indirect detection of dark matter using neutrino telescopes, on the other hand, relies on
WIMPs being efficiently captured in the Sun, where they then annihilate and generate high-energy
neutrinos. KK dark matter becomes captured in the Sun most efficiently through its spin-dependent
scattering off of protons. Since this cross section scales with $1/m^4_{B^{(1)}}$, the constraints
presented in this article somewhat limit the rates which might be observed by next generation
high-energy neutrino telescopes, such as IceCube \cite{icecube}. In particular, a 800 GeV LKP
could generate approximately 20 or 3 events per year at IceCube for KK quark masses 10\% or 20\%
larger than the LKP mass, respectively \cite{neutrinokribs}. Over several years of observation, a
rate in this range could potentially be distinguished from the atmospheric neutrino background. Larger
volume experiments (multi-cubic kilometer) would be needed to detect an LKP which was significantly
heavier than $\sim$~TeV.

\section{Conclusions}

In this paper we have re-investigated the bounds on the compactification scale of
Universal Extra Dimension extension arising from electroweak precision observables
measured at LEP1 and LEP2. 
The lower bound is improved to be $M_c\equiv
R^{-1}>700$ $(800)$ GeV at the 99\% (95\%) confidence level.
There are two origins for the improvement of the bound compared to
\cite{Appelquist:2002wb}.  First, when taking two-loop Standard Model
contributions to the electroweak precision parameters into account, the
KK-contributions to the $\hat{T}$ parameter no longer cancel against
Higgs-dependent contributions in the heavy-Higgs-mass limit.  This
differs from the situation of Ref.\cite{Appelquist:2002wb} where only
one-loop Higgs-dependent contributions were taken in to account.  This
lack of cancellation is the reason for elimination of the heavy-Higgs-mass
and low $M_c$ region.   Second, the inclusion of LEP2
data into the analysis necessitated the use of an extended set of well-determined
electroweak precision observables as shown in Ref.\cite{Barbieri:2004qk}. 
These new EWPO provide additional constraints, further lifting the bound on $M_c$.

The new constraint presented in this article can have a significant impact on the phenomenology of Kaluza-Klein dark matter. Indirect detection techniques often rely on the efficient annihilation of dark matter particles in the galactic center, galactic halo, or in dark substructure. The models with the highest annihilation rates of Kaluza-Klein dark matter are those with a low compactification scale, and are thus excluded by the results of this study. The prospects for direct detection are also somewhat reduced by this constraint.

\section{Acknowledgements}
We are grateful to Riccardo Rattazzi and Alessandro Strumia for helpful communications. JMR and TF would
especially like to thank the Department of Physics of the University of California, Davis,
where much of this work was done, for their kind hospitality and for support from the
HEFTI visitors program.  The work of TF was supported by ``Evangelisches Studienwerk Villigst e.V.'' and
PPARC Grant No. PPA/S/S/2002/03540A. DH is supported by the US Department of Energy and by NASA grant NAG5-10842.
This work was also supported by the `Quest for Unification' network,  MRTN 2004-503369.

\newpage

\appendix

\section{NDA for Universal Extra Dimensions}

Here we outline the na\"{\i}ve dimensional
analysis (NDA) of the UED contributions to the EWPO.
The 4D effective action of a universal EW theory is
\cite{Barbieri:2004qk}
\bea\label{4Dbarbieriaction}
  \mathcal{L}_{4eff}=&-&\frac{1}{4}b_{\mu\nu}b^{\mu\nu}
    -\frac{1}{4}w^I_{\mu\nu}w^{I\mu\nu}+(D_{\mu}h)^{\dagger}D^{\mu}h\nonumber\\
    &+&\frac{1}{v^2}(c_{wb}O_{wb}+c_hO_h+c_{ww}O_{ww}+c_{bb}O_{bb})+...,
\eea
where the lower case fields denote the 4D effective fields,
$v=174$ GeV is the Higgs VEV, the operators $O$ are defined below
and the parentheses contain Higgs potential and Yukawa
terms.
${O_{wb},O_h,O_{ww},O_{bb}}$ form a basis of the universal
dimension 6 operators
\cite{Barbieri:2004qk,Barbieri:1999tm}. All other universal dimension
6 operators are equivalent to the ones given via the equations of motion.  

Starting from the 5D theory and KK-expanding the fields, the only
operators which in the 4D effective theory lead to dimension 6
operators which solely depend on light fields (zero-modes) are the
5D analogues of the operators $O_{wb},O_h,O_{ww},O_{bb}$
\bea
  O_{WB} &\equiv & H^{\dagger}T^I H W^I_{MN}B^{MN}\nonumber\\
  O_H &\equiv &|H^{\dagger}D_MH|^2\nonumber\\
  O_{BB}&\equiv &\frac{1}{2}(\partial_R B_{MN})^2\\
  O_{WW}&\equiv &\frac{1}{2}(D_R W_{MN})^2\nonumber
\eea
and operators which are equivalent to them via the 5D equations of
motion.  The parameters $\hat{S},\hat{T},Y,W$ are related to the operator
coefficients by \cite{Barbieri:2004qk}
\bea\label{STYWbarbdef}
  \hat{S}&=&\frac{2}{\tan\theta_w}c_{wb}\nonumber\\
  \hat{T}&= &-c_h\nonumber\\
  W&= &-g^2c_{ww}\\
  Y&= &-g^2c_{bb}.\nonumber
\eea

Na\"{\i}ve dimensional analysis of the 5D action
yields\footnote{Note that NDA numerical factor differs \cite{Papucci:2004ip} from the usually quoted $24\pi^3$ of
Ref.\cite{Chacko:1999hg} for a 5D theory.  We thank Riccardo Rattazzi for discussions on this
point.}
\bea\label{5Dbarbieriaction}
  \mathcal{L}_{5D}=&\frac{\Lambda^5 \pi R}{24 \pi^2}\biggl[-\frac{1}{4\Lambda^2}B_{MN}B^{MN}
    -\frac{1}{4\Lambda^2}W^I_{MN}W^{IMN}+\frac{1}{\Lambda^2}(D_M H)^{\dagger}D^M H  \nonumber\\
    &+ \frac{1}{\Lambda^2}O_{WB}+ \frac{1}{\Lambda^2}O_H
    + \frac{1}{\Lambda^4}O_{WW}+\frac{1}{\Lambda^4}O_{BB} \biggr]+\ldots,
\eea
where here $\Lambda$ is the strong coupling scale of the 5D theory.
After evaluating this action on the zero modes, integrating over the extra dimension,
and then canonically normalizing the fields, the 4D effective action for the light fields
reads
\bea\label{4DNDAaction}
  \mathcal{L}_{4eff}=&-&\frac{1}{4}b_{\mu\nu}b^{\mu\nu}+(D_{\mu}h)^{\dagger}D^{\mu}h
    -\frac{1}{4}w^I_{\mu\nu}w^{I\mu\nu}\nonumber\\
    &+&\frac{24\pi^2}{\pi R\Lambda^3}O_{wb}+\frac{24\pi^2}{\pi R\Lambda^3}O_h+\frac{1}{\Lambda^2}O_{ww}
	+\frac{1}{\Lambda^2}O_{bb}+\ldots,
\eea
together with the NDA expression for the largest gauge coupling of the theory -- the QCD
coupling $g_3$ (a related expression
holds for the largest Yukawa coupling if it becomes strong at the same scale $\Lambda$)
\beq\label{NDAgauge}
g_3^2 \sim \frac{24\pi^2}{\pi R \Lambda} .
\eeq

Further, matching the NDA action Eq.(\ref{4DNDAaction}) to Eq.(\ref{4Dbarbieriaction}) and
using Eq.(\ref{STYWbarbdef}) leads to the NDA estimates for the contributions to the
leading EWPO
\bea\label{STYWfromNDA}
  \hat{S}&\sim& \frac{48\pi v^2}{\tan\theta_w R\Lambda^3}\nonumber\\
  \hat{T}&\sim& \frac{24\pi v^2}{R \Lambda^3}\nonumber\\
  W&\sim&\frac{ g^2 v^2}{\Lambda^2}\\
  Y&\sim&\frac{ g^2 v^2}{\Lambda^2}.\nonumber
\eea

Given the measured size of the QCD coupling at $m_Z$, the NDA expression for the
largest gauge coupling Eq.(\ref{NDAgauge}) leads to a limit on the ratio of the
cutoff $\Lambda$ to the KK mass scale $M_c =1/R$ given by
\beq
R\Lambda \lsim 48
\eeq
If we assume this estimate of the bound on the cutoff scale is saturated
then from the NDA estimates of the EWPO, we expect $W\sim Y \sim 0.1 \hat{T} \sim
0.03 \hat{S} \sim 10^{-4} (v R)^2$

\end{document}